\documentclass[aps,prl,twocolumn,show]{revtex4}
\pdfoutput=1
\usepackage{epsfig}
\usepackage{amsmath}
\usepackage{amssymb}
\usepackage{color}

\begin{document}

\title{Properties of Cooperatively Induced Phases in Sensing Models}
\author{Stuart A. Sevier,$^*$ }
\affiliation{Department of Physics and Astronomy University of California, Los Angeles, CA 90095, USA \\ Center for Theoretical Biological Physics, Rice University, Houston, TX 77005}

\author{Herbert Levine }
\affiliation{Department of Bioengineering, Center for Theoretical Biological Physics, Rice University, Houston, TX 77005}
\date{\today}

\begin{abstract}
A large number of eukaryotic cells are able to directly detect external chemical gradients with great accuracy and the ultimate limit to their sensitivity has been a topic of debate for many years.   Previous work has been done to understand many aspects of this process but little attention has been paid to the possibility of emergent sensing states.  Here we examine how cooperation between sensors existing in a two dimensional network, as they do on the cell's surface, can both enhance and fundamentally alter the response of the cell to a spatially varying signal.   We show that weakly interacting sensors linearly amplify the cell's response to an external gradient while a network of strongly interacting sensors form a collective non-linear response with two separate domains of active and inactive sensors forming what have called a "1/2-state" .  In our analysis we examine the cell's ability to sense the direction of a signal  and pay special attention to the substantially different behavior realized in the strongly interacting regime.  \end{abstract}
\pacs{5,87}

\maketitle

\section{Introduction}
Cells of all types move under the influence of chemical signals in order to participate in important biological functions \cite{Parent1999} . To do this cells have developed special sensors which bind to particular molecules which make up an external signal.  Small cells commonly employ a integration process where the cell compares concentration signals over time along their path \cite{Segall1986} . Large cells, on the other hand, can directly measure concentration gradients across their cell bodies and do not have to integrate the signal as they travel along \cite{Arkowitz1999, Samadani2006b}.  In both schemes the cell is able to sense the extremely small differences in the direction and magnitude of the chemical gradient.  In the case of eukaryotic spatial sensing the cell is sensitive to a $1\%-2\%$ difference in concentration across the cell body \cite{Song2006f} .  This exquisite ability is especially impressive considering the inherently noisy and dynamic nature of the sensor and signal.

There have been numerous theoretical attempts at  understanding this sensitivity and the limits to it, starting first with the seminal work completed by Berg and Purcell \cite{Berg1977} which posited a minimal uncertainty in concentration sensing due to the diffusion of the signal itself. Further work has been done to include the effects of ligand-sensor dynamics and cooperativity  on sensing \cite{Bray1998,Mello2003,Mello2004b,Bialek2005b,Keymer2006b,Wang2007,Bialek2008,Endres2009,Skoge2006} as well as more recent works seeking to understand the uncertainty associated with spatial sensing \cite{Hu2010a,Hu2010}. The models and ideas have had many successes but some questions and undiscovered possibilities still remain.
\begin{figure}[b]
\includegraphics[width=0.7\linewidth,clip=]{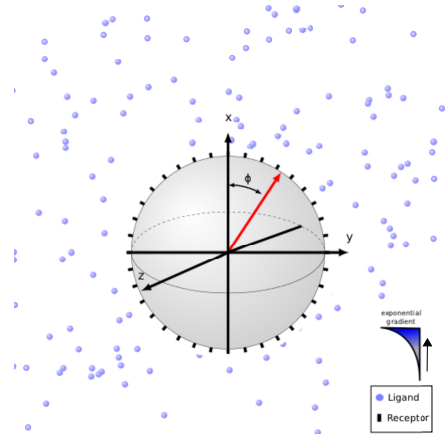}
 \caption{\label{fig:lattice} (color online) A schematic of the cell in a chemical gradient. }
 \end{figure}

One notable question regards the existence and possible benefits of interacting sensors.   Sensors in bacteria are arranged in tight clusters but the theoretical debate over whether or not their interactions enhance sensing is still active.  One intriguing model of cooperativity maps the activity of sensors onto an interacting Ising model where it is clear that the interactions can enhance the response of the sensors to the direction and concentration of an external stimulus \cite{Hu2010}.  It is however not clear that the signal-to-noise ratio is increased for interacting sensors if they the integrate the signal \cite{Skoge2011b} where the precise dynamic ansatz \cite{Sun2014} and non-equilibrium nature \cite{Skoge2013} of the model can change the effects of cooperativity on the signal-to-noise ratio; in some instances improving it and others destroying it.   The issues present for dynamic integration may not be that important for eukaryotic sensing,  which can occurs even in stationary cells \cite{Samadani2006b}.  Direct evidence for interacting sensors in eukaryotic systems is hard to find, but has not be ruled out either.

In this letter we will not address the precise biological mechanisms which generate interactions between sensors, but will instead suppose their existence and assume that the sensors are evenly distributed over the cell and that they interact locally with one another.  Here we will be especially interested in the additional phenomena exhibited by the sensors when they are placed in a true 2D network, as exists on the surface of the cell.  In doing so we will uncover two regimes of cooperative sensing which are separated by a critical value in interaction strength.  For weakly or non-interacting sensor the system displays linear behavior in which interactions quantitatively enhance, but do not qualitatively alter, the performance of the cell's sensing.  This regime is identical in most ways to the behavior seen in 1D systems \cite{Hu2010} .   For strongly interacting sensors the system can form a collective response which is qualitatively different behavior than the linear phase and what is predicted for a strictly 1D system.  This behavior has been unexplored in previous works.
 \begin{figure}[t]
\includegraphics[width=.9\linewidth,clip=]{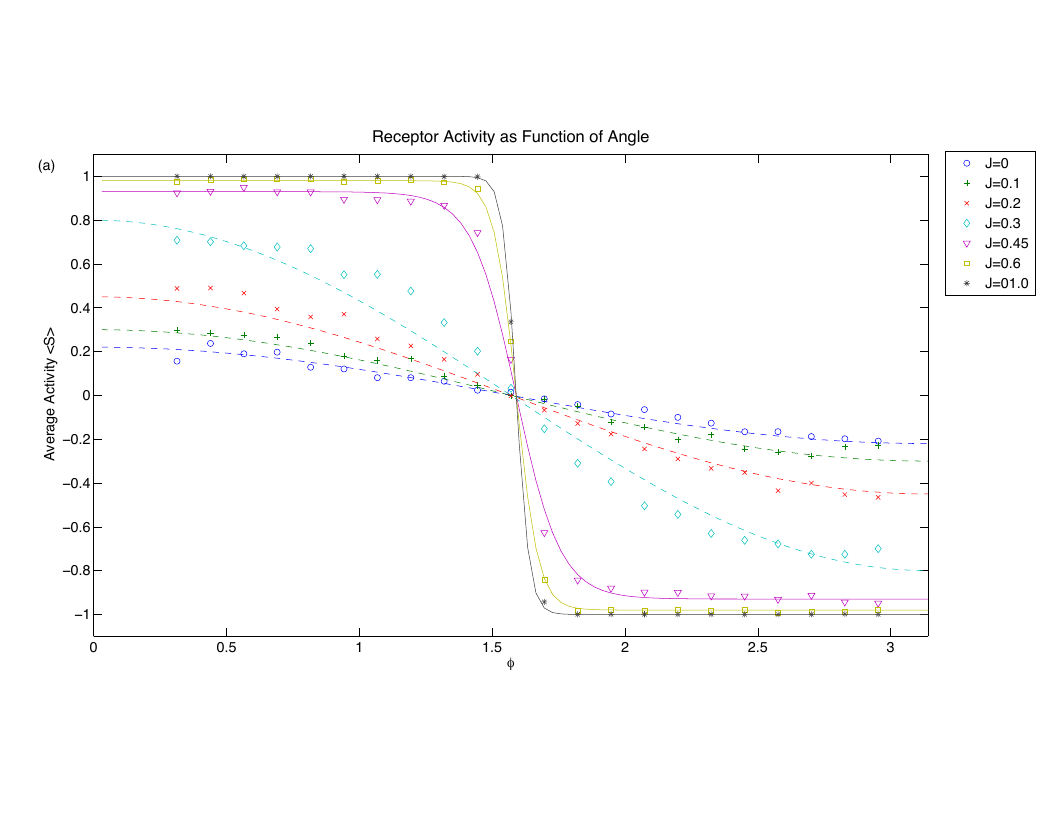}
\includegraphics[width=.9\linewidth,clip=]{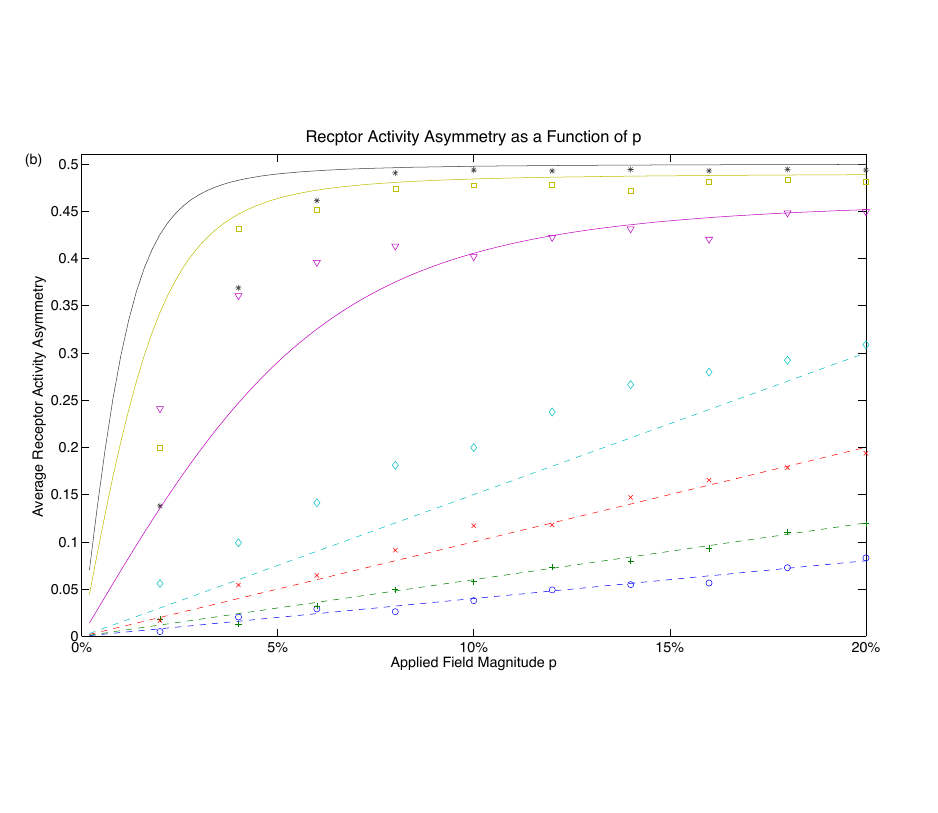}
 \caption{\label{fig:lattice} (color online) The figures were made by placing 5000 sensors in a spatially varying field with strength p at $H_{0}=0$. Figure 2a shows the average response of the sensors $<S(\phi)>$ in the plane of an applied external field from $\phi=0:\pi$ at a fixed gradient steepness of $p=10 \% $  . Figure 2b shows the average response asymmetry along the applied gradient direction (normalized by the number of sensors $N$) as a function of the applied gradient strength $p$.  In both figures the points are from Monte-Carlo simulations while the dotted curves correspond to the linear-response $<S_L(\phi)>$  and the solid curves to the collective "1/2-state" response $<S_C(\phi)>.   $}
 \end{figure}

\subsection{Model}
To explore the consequences of cooperativity we will consider a large number $N$ of sensors evenly distributed \cite{Cai2011} over the sphere where each sensor is able to take on two states: active ($S_{i}=+1$) or inactive ($S_{i}=-1$) .  We will imagine there are two effects felt by a single sensor: the first is a nearest-neighbor interaction energy $J$ (which could be the result of sensor-sensor interactions as well as internal 'downstream' dynamics) in units of thermal energy which acts only between nearest neighbor sensors $<i,j>$  and the second an energy $H_{i}$ associated with a single active state (or inactive state $-H_{i}$) due to an external chemical field. The total energy of the system thus given by
\begin{equation}
\label{eq:H_1}
E=-J\sum_{<i,j>}S_{i}S_{j}-\sum_{i}H_{i}S_{j}
\end{equation}
This form is precisely that of an interacting Ising model with nearest-neighbor interactions. The energy $H_{i}$  is the external input and is derived from matching the Boltzmann probability $P_{on}(i)=e^{H_{i}}/(e^{H_{i}}+e^{-H_{i}})$ to the sensor-ligand kinetic probability $P_{on}(i)=C_{i}/(C_{i}+K_{d})$ in chemical equilibrium where $K_{d}=k_{-}/k_{+}$ is the dissociation constant. Comparing the two gives a single-site free energy 
\begin{equation}
\label{eq:H_1}
H_{i}=\frac{1}{2}ln\frac{C_{i}}{K_{d}}
\end{equation}
We will view this local free-energy as a external field which effects the sensor energy at each site as $H_{i}S_{i}$.  

To explore the properties of the system we shall specify the exact form of the external gradient $H_{i}$, and a particularly nice choice is a symmetric exponentially disturbed concentration pointed at an angle $\hat{\varphi} $ in the $x-y$ plane  given by  $C_{i}=C_{o}Exp[2p cos\left(\phi_{i}-\hat{\varphi}\right)]$ .  We have made this choice because it causes the local field $H_{i}$ to break down nicely into a constant $H_{0}=\frac{1}{2}ln\frac{C_{0}}{K_{d}}$ and spatially varying component $H\left(\phi_{i}\right)=p cos\left(\phi_{i}-\hat{\varphi}\right)$ making the properties we wish to examine easy to explore.

\subsection{Results}
The existence of collective states for interaction strengths over a critical coupling $J_{c}$ is a signature of 2D systems and this feature makes it necessary to  employ different strategies for a weakly interacting ($J<J_{c}$) systems, which respond in a linear fashion, than for strongly interacting  ($J>J_{c}$) systems. The latter can form a collective state, which we will call the "1/2-state" and fall outside of a linear response analysis.  

For weakly interacting sensors in a gradient where $H_{0} \sim 0$ we can employ the power of linear response theory, using the 2D Ising zero field response function, to compute the linear response of the sensors $<S_{L}>$  when placed into a weak spatially varying field.  The response is given as function of angle in the applied field plane as
\begin{eqnarray*}
<S_{L}(\phi)> & = & p \frac{N}{3}c_{1}(J) \cos(\phi -\hat{\varphi})
 \end{eqnarray*}
where the coefficient $c_{1}(J)$ depends on the interaction strength $J$ and is the only coefficient left over from the spherical harmonic expansion of the response function (see appendix).   Figure 2 compares the analytical expression to results from Monte-Carlo simulations of interacting sensors with  the predicted linear-response given by the dotted lines.  Just as in one dimension the linear response of the sensor activity to the external gradient for the 2D system with $J<J_{c}$ is pointed in the direction of the field and proportional to the locally applied field strength $p$ with increased response $c_{1}(J)$ for $J>0$. 
 
 The cell can use the response of the sensors to construct an estimate for the gradient direction by considering the spatial dependence of the average activity.  For small $p$ the values are normally distributed and can be used as estimators for the external gradient direction and magnitude.  The estimators provide a maximum-likelihood estimation (MLE) for the lower-bound \cite{Kay1993} in the uncertainty of the cell's estimate of the direction of the gradient yielding a minimum of $\sigma_{\hat{\varphi}}^{-2} =N\pi^{2}c_{1}(J) p^{2}/12$ which is similar to the previously obtained 1D result  \cite{Hu2010}.

In the strongly interacting regime  ( $J>J_c$ ), when the constant field $|H_{0}|$ is large, the system simply freezes into a totally active or inactive state and is not capable of spatial sensing.  However, when the constant field is smaller than the asymmetric component $p>|H_{0}|$ we have found that the response of strongly interacting sensors is to form a robust collective state with two regions of active and inactive sensors.  When $H_{0} \sim 0$ the domain perfectly divides the cell into active and inactive regions -forming a "1/2-state"- separated along the equator of the cell orthogonal to the direction of the external field.  

The collective nature of the strongly interacting state requires a different approach than previously considered.  Following the Landau-Ginzburg prescription for collective behavior one can construct the strongly interacting sensor profile (see appendix) as
\begin{equation}
\label{eq:H_1}
<S_{C}(\phi) >=S_{0}tanh(A(p,J)cos(\phi-\hat{b}))
\end{equation}
The profile shows strongest response in the direction $\hat{b}$ (which can be different than the applied field direction $\hat{\varphi}$ ) and smoothly crosses over from active to inactive.  The coefficient $A(p,J)$ determines how sharp the transition from active to inactive is and $S_{0}$ the average activity in the two domains away from the transition region.   The analytical expression for the response of the collective state for various interaction strengths $J>J_{c}$ are shown by solid curves in figure 2 and are compared to Monte-Carlo simulations with corresponding interaction strengths .
 
The collective profile $<S_{C}(\phi) >$ is near the maximum possible response for a system of sensors in a spatially varying field where there is an instantaneous transition in sensor activity from on to off, separating the sphere into two completely active and inactive domains of equal size, something we have called a "1/2-state".  The most favorable configuration of the "1/2-state" is complete alignment with the external field, however it will fluctuate away from perfect alignment and encounter an energy cost due to the strain created between the sensors and the applied field.  This energy is given as a function of the separation $\gamma=\hat{\varphi}-\hat{b}$ between the direction of the external field $\hat{\varphi}$ and the direction of the "1/2-state" $\hat{b}$ as (see appendix) 
\begin{equation}
\label{eq:H_1}
E=- \sum_{i} H_{i} S_{i}=- \frac{p N S_{0}f(A)}{2}cos\gamma
\end{equation}
In the small deviation limit $\gamma \approx 0$ the Boltzmann energy probability distribution for the alignment of the "1/2-state" is a normal distribution given by 
\begin{equation}
\label{eq:H_1}
G(\hat{b}|\hat{\varphi})=\frac{1}{\sqrt{2\pi\sigma^{2}}}e^{-\frac{(\hat{\varphi}-\hat{b})^2}{2\sigma^{2}}}
\end{equation}
with mean  $\hat{\varphi}$ and variance $\sigma^{-2}=pNS_{0}f(A)/2$. From the probability density function $G$ and the principles of MLE  \cite{Kay1993} we can calculate the expected minimum knowledge of the external gradient direction $\hat{\varphi}$ setting the theoretical lower limit of the cell's ability to sense the gradient direction at any given instant given as
\begin{equation}
\label{eq:H_1}
\sigma_{\hat{\varphi}}^{2}=\frac{2}{pNS_{0}f(A)}
\end{equation}

When the transition from the domain of active sensors to inactive sensors is instantaneous $f(A) \rightarrow 1,\;S_{0}\rightarrow1$ (which happens for modest choices of $J,\;p$) the limit to the uncertainty in direction sensing is set by the number of sensors and the applied field magnitude $\sigma_{\hat{\varphi}}^{-2} \sim pN$. This scaling is independent of the particular choices made for the strength or nature of the interaction between sensors.   Figure 3a compares Monte-Carlo data to the analytical expression for  the uncertainty in direction sensing at $H_{0} \sim 0$ for  both non-interacting, weakly interacting and strongly interacting systems. 

The expressions presented so far are only true for a purely spatially varying field situated at $H_{0} \sim 0$ which corresponds to the area where $C_{0} \sim K_{d}$ .  Away from this region most of the qualitative results will remain true with modified quantitative expressions.   For $J>J_{c}$ a collective state can still form for small $H_{0}$ but with unequally sized domains separated by the point where $H_{0}+pcos(\phi)=0$ which occurs at the angle $\phi=cos^{-1}(-\frac{H_{0}}{p})$ .  This changes the energy cost of the system being misaligned from the external field resulting in a decreased accuracy in determining the direction of the gradient with variance given for the idealized "1/2-state" in a non-zero constant field by $\sigma_{\hat{\varphi}}^{-2}=p\sqrt{1-\frac{H_{0}^{2}}{p^{2}}}N/2$ .   Figure 3b shows the uncertainty in direction for non-zero background fields with fixed asymmetry magnitude $p$.  The noticeable deviation of the simulation data from the analytical expression as $|H_{o}| \rightarrow p$ for strongly interacting sensors is due to the extra source of fluctuations for increasingly small domain sizes.  As the region of active (or inactive) sensors becomes smaller the fluctuation in the domain itself become significant and the likelihood that a fluctuation can push the system of sensors into a completely active (or inactive) state becomes large resulting in a decreased ability to sense the gradient direction. 
\begin{figure}[t]
\includegraphics[width=.9\linewidth,clip=]{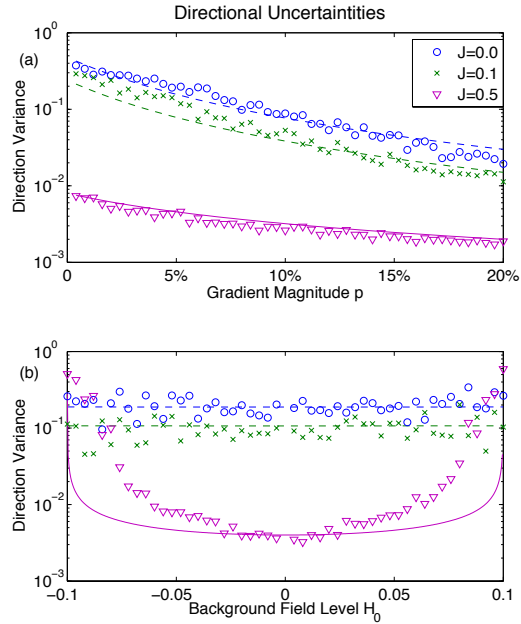}
 \caption{\label{fig:lattice} (color online) Figure 3a is the uncertainty $\sigma_{\varphi}$ in direction for changing values of asymmetry magnitude $p$ at $H_{0} \sim 0 $. Figure 3b  is the uncertainty $\sigma_{\varphi}$ in direction for varying background field levels $H_{0}=1/2 ln\frac{C_{0}}{K_{d}}$ at a fixed asymmetry magnitude of $p=10 \%$ .  In both figures the circles and dotted line correspond to the Monte-Carlo and analytical results for a non-interaction and weakly-interacting systems while the squares  and solid lines show the same, respectively, for a strongly-interacting "1/2-state".  
 }
 \end{figure}

\subsection{Discussion}

The existence of the "1/2-state" for strongly interacting sensors is a novel response of sensing models to external cues resulting in an enhanced ability for the cell to determine the gradient direction when $H_{0} \sim 0$. 

For all 1D, as well as weakly and strongly interacting 2D systems, the lower-bound decreases with increased sensor number as $\sigma_{\varphi}^{2} \sim N^{-1}$. However there is a substantial difference in the gradient dependence, where for both 1D and 2D weakly interacting systems $\sigma_{\varphi}^{2} \sim p^{-2}$  whereas for strongly interacting 2D systems $\sigma_{\varphi}^{2} \sim p^{-1}$  .  This difference  means that the "1/2-state" is significantly better at directional sensing for small $p$ by an order of magnitude (see fig. 2a)

One issue that needs further study is the rate at which the sensing apparatus can adapt to changing stimuli if it is operating in the strongly-interacting regime. For strongly-interacting systems with constant field components $p<H_{0}$ the favorable configuration of sensors is to be totally active or inactive. Because of this for $J>J_{c}$ the history of the environment plays a crucial role in determining the time-dependent collective response of the system. This might add a significant challenge for the cell to use the strongly interacting approach to sense its surroundings, though adaptive changes in  the dissociation $K_{d}$ or coupling $J$ could act to protect the system from hysteresis. 

 We would like to emphasize that the particular model used to study the effects interacting sensors should be thought of as an avenue to understanding more general mechanisms of sensing.  In this letter we have occupied ourselves with the examination of both an idealized domain formation, which we have called a "1/2-state", and a realization of such a state generated by a locally interacting Ising model. Numerous interacting statistical models could be capable of generating the collective behavior necessary for the formation of a domain wall and the analysis concerning the idealized "1/2-state" only requires that there is some mechanism for sensor coordination.  Once formed, the properties of the collective state are only effected by the energy associated with an individual sensor, making the concept of a "1/2-state" extend beyond Ising like models of sensors. 
 
While the evidence for strongly interacting sensors in eukaryotic systems is currently lacking it is clear that theoretically their existence would dramatically increase the cell's ability to accurately respond to external cues in a number of scenarios.    Our analysis clearly demonstrates that interacting sensors can not only increase the ability of the cell to response to external cues, but can fundamentally alter the nature of the response resulting in drastically increased accuracy.

\begin{acknowledgments}
This work was supported by the U.S. National Institutes of Health (PO1 GM078586). Support was also obtained from the National Science Foundation Physics Frontier Center program grant no. PHY-1427654. S.S. was also partially supported by the National Science Foundation Graduate Research Fellowship under Grant No. DGE-1144087.  S.S. would like to dedicate his portion of the work to the memory and encouragement of his grandfather, N.E. Abbott.  

\end{acknowledgments}

%I want to pull in all the citations so that they're easier to reference if needed .  \cite{Aquino2010,Berg1977, Berg1977a, Bialek2008, Bialek2005b, Bray1998, Deng2003b, Dusenbery1998, Endres2008c, Endres2009, Endres2009a, Ferraro2008c, Fuller2010c, Fuller2010, Gamba2007, Hu2010, Hu2011c, Hu2010a, Iglesias2008b, Jilkine2011, Keymer2006b, Macnab1972, Mello2004b, Mello2003, Mora2010c, Parent1999, Rappel2008, Rappel2008d, Samadani2006b, Segall1986, Shi1998b, Skoge2006b, Skoge2006, Skoge2011b, Skoge2013, Smallshaw2006, Song2006f, Sourjik2004b, Sourjik2002, Sun2014a, Sun2014, Thar2003b, Ueda2001, Ueda2007b}

\bibliographystyle{unsrt}
\bibliography{sensing}

 \section{Appendix}

\subsection{Linear Response}
For non-interacting sensors the 2D nature of the network does not substantially effect any of the results from 1D systems.  To see this we will construct the systems energy  and factorize it into direct products of the response activity  $M_{1,2}$  as 
\begin{equation}
\label{eq:H_1}
E=-\sum_{i}H_{i}S_{j}=-MH_{0}-\left(M_{1} \alpha_{1}+M_{2}\alpha_{2} \right)
\end{equation}
where the sum is done over all $N$ sensors and we have used $(\alpha_{1},\alpha_{2})=p(cos\hat{\varphi},sin\hat{\varphi})$ and $M=\sum_{n}S_n,\;M_{1}=\sum_{n}S_n cos \phi_n,\;M_{2}=\sum_{n}S_n sin \phi_n$ and have factored the field as  $H_{i}=H_{0}+H(\phi_{i})$ and taken a nice form for the spatial component $H\left(\phi_{i}\right)=pcos\left(\phi_{i}-\hat{\varphi}\right)$ and the constant component $H_{o}=\frac{1}{2}ln\frac{C_{0}}{K_{d}}$. The rationale and matching of the field components to the binding probability is given in the main text.  Since the energy has this simple form the expected values are easy to calculate and are given by 
\begin{equation}
\label{eq:var}
<M_{1,2}>=2\frac{\partial lnZ }{\partial \alpha_{1,2}}=\mu\alpha_{1,2}
\end{equation}
where $Z$ is the partition function for the system and  $\mu = N \frac {C_{0} K_{d}}{4(C_{0}+K_{d})^{2}}$ .  The variance for the two quantities can be calculated with ease due to the structure of the energy and is given by 
\begin{equation}
\label{eq:var}
Var(M_{1,2})=2\frac{\partial <M_{1,2}>}{\partial \alpha_{1,2}} =2 \mu
\end{equation}

The linear response of the sensors $<S_{n}>$  for $J_{c}>J > 0$ when placed into the applied field is given by 
 \begin{equation}
\label{eq:H_1}
<S_{n}>=\sum_{m}\chi_{n,m}H\left(\phi_{m}\right)
\end{equation}
Though the 2D response function $\chi(J)_{m,n}$ is known for planar or toroidal geometries \cite{Baxter2007,Pathria1996} the response $\chi(J)_{m,n}$ is only sensitive to the geometry  (i.e. scalar distance) of the system and are unchanged by the global topology.  Indeed we have found a critical value of $J_{c}\sim.44$ for our 2D spherical network matching the known results for other 2D systems.  This will allow us to freely use $\chi(J)_{m,n}$ in our spherical topology however only in the regime $J<J_{c}$.   We will write the response function, which only depends on the length between the two points on the sphere $r_{n,m}=|\vec{r}_{n}-\vec{r}_{m}|=\sqrt{2}R\sqrt{1-\hat{r}\cdot\hat{r}'}=\sqrt{2}R\sqrt{1-cos\Delta}$, in a spherical harmonic basis as
\begin{eqnarray*}
\chi(\theta,\phi;\theta',\phi')& = & \sum_{l}c_{l}(J)P_{l}\left[cos\Delta \right]\\
 & = & \sum_{l} \frac{4\pi}{2l+1}\sum_{l=-m}^{l=+m}c_{l}(J)Y_{lm}\left(\theta,\phi\right)Y_{lm}^{*}\left(\theta',\phi'\right)
 \end{eqnarray*}
where we have related the Legendre functions $P_{l}$ to the spherical harmonic functions $Y_{lm}$ through the Addition Theorem of Spherical Harmonics \cite{jackson_classical_1999} and the coefficients are calculated by projecting the response function \cite{Baxter2007, Pathria1996} onto Legendre functions. 
 \begin{equation}
\label{eq:corr}
\chi(J)_{m,n}=\frac{\xi^{3/4}}{2^{21/8}\pi} \frac{e^{-r_{n,m}/ \xi}}{r_{n,m}^{2}}
\end{equation}
where $\xi=\frac{1}{4(J-J_{c})}$ is the correlation length near the transition. The use of the above function has multiple potential issues concerning its validity in the short range limit as well the Euclidian versus arc length distance between points along the sphere.  We have found neither to be too concerning on theoretical or numerical grounds.  

To evaluate the expression we convert the summation into an integral by assuming an even density of many sensors across the surface of the sphere by including a factor of $\frac{N}{4{\pi}^{2}}$  and compute the response \textit{in the frame of the external field}  then move back into the frame of the cell.  The average response is given as 
\begin{eqnarray*}
<S(\tilde{\theta})> & = & \frac{N}{4\pi}\int d\Omega'\chi(\tilde{\theta},\tilde{\phi};\theta',\phi')H(\theta',\phi')\\
 & = & p\frac{N}{4\pi}\sum_{l,m} \frac{4\pi}{2l+1} c_{l}Y_{lm}\int d\Omega'Y_{lm}^{*}\left(\theta',\phi'\right)cos\theta'\\
 & = & p\frac{N}{4\pi}\sum_{l,m} \frac{4\pi}{2l+1} c_{l}Y_{lm}\int d\Omega'Y_{lm}^{*}\left(\theta',\phi'\right) \sqrt{\frac{4\pi}{3}}Y_{10}\\
 & = & p\frac{N}{3}c_{1}cos\tilde{\theta}
\end{eqnarray*}
where only the first coefficient $c_{1}(J)$ is needed due to the symmetry created by our choice in the external field. Moving from the frame of the applied field back to the frame of the cell gives the average response as 
 \begin{equation}
<S_{L}(\phi)>=p\frac{N}{3}c_{1}(J) cos(\phi-\hat{\varphi})
\end{equation}

Now we can calculate the expected values of $M_{1,2}$  by averaging over the system as 
\begin{eqnarray*}
<M_{1}> & = &\int d\Omega<S(\theta,\phi)>cos\phi sin\theta\\
 & = & pN \frac{\pi^{2}}{6}c_{1}(J)cos\hat{\varphi}=\mu(J)\alpha_{1}\\
\end{eqnarray*}
and
\begin{eqnarray*}
<M_{2}> & = & \int d\Omega<S(\theta,\phi)>sin\phi sin\theta\\
 & = & pN \frac{\pi^{2}}{6}c_{1}(J)sin\hat{\varphi}=\mu(J)\alpha_{2}\\
\end{eqnarray*}
where we have used $(\alpha_{1},\alpha_{2})=p(Cos\hat{\varphi},Sin\hat{\varphi})$ and $\mu(J)=N \frac{\pi^{2}}{6}c_{1}(J)$. The coefficient $c_{1}(J)$ is plotted below as a function of $J$. 
 \begin{figure}[h]
\includegraphics[width=.8\linewidth,clip=]{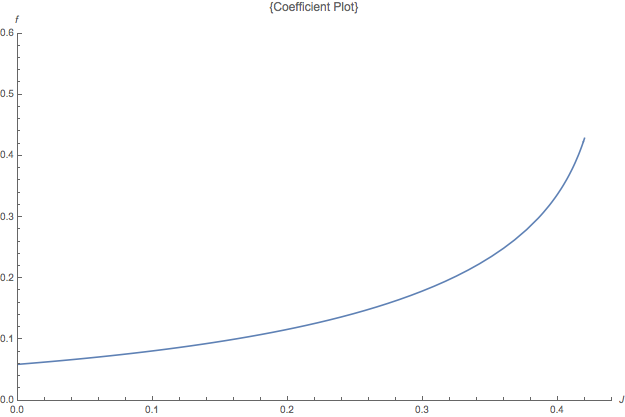}
 \caption{\label{fig:lattice} Legendre coefficient $c_{1}(J)$ for response function expansion as a function of interaction strength $J>H_{c}$ }
 \end{figure}

The calculation of the variances $Var(M_{1,2})$ is made easy again from the energy structure of the system (even in the presence of interactions) and is given by eq\ref{eq:var}. To find the lower-bound on the accuracy of the cell to sense the direction of the gradient let us consider the probability density function of the two estimators $M_{1,2}$ 
\begin{eqnarray*}
G = \frac{1}{2\pi \sigma^{2}}Exp(-\frac{(M_{1}-<M_{1}>)^{2}+(M_{2}-<M_{2}>)^{2}}{2\sigma^{2}})
\end{eqnarray*}
which is the probability density of finding values $M_{1,2}$ given an applied field with values $\alpha_{1,2}$ where $\sigma^2=Var(M_{1,2})$.  With $M_{1,2}$ as estimators we can use the properties of maximum-likelihood estimation \cite{Kay1993} to calculate the Cramer-Rao lower bound associated with the direction of the applied field $\hat{\varphi}$
\begin{eqnarray*}
\sigma^{2}_{\hat{\varphi}} & = & \frac{2}{\mu(J) p^{2}} =\frac{12}{N \pi^{2} c_{1}(J) p^{2}} 
\end{eqnarray*}

\subsection{Collective State}

The collective state created by the strongly interacting sensors is a division of the sphere into two regions, one with average activity and one with average inactivity.  In the main text we have referred to this as a "1/2-state"  because roughly half of the sensors are active and half are inactive and the spatial projections $M_{1,2}$ approach their maximum value of $1/2$ . The average sensor activity profile can be determined through the Ginzburg-Landau free energy equation of motion for the average activity 
\begin{equation}
\label{eq:H_1}
k\nabla^{2}S(x)=H(x)-aS(x)-bS(x)^3
\end{equation}
where the coefficients $k, a, b$ are well defined functions and constants of the coupling $J$.   The homogenous solution $(H=0)$ to this equation with boundary conditions of $S(x>>0)\rightarrow S_{0}$ and$S(x<<0)\rightarrow -S_{0}$ is 
\begin{equation}
\label{eq:H_1}
S_{C}(\phi)=S_{0} \ tanh(Ax)
\end{equation}
with $x= cos\phi$.
To generate an approximate solution to the inhomogeneous equation $H(\phi)=p cos\phi$ we can expand the homogeneous solution near $S(\phi)=0$ in the inhomogeneous equation generating a matching condition which yields $A \sim \frac{p}{k+a}$, while the boundary conditions $S(x>>0)\rightarrow S_{0}$ and$S(x<<0)\rightarrow -S_{0}$ are matched to the system by considering the self-consistent average sensor activity at the boundary given by $S_{0}=tanh(JS_{0}+H(\phi=0))$.

Since the sloshing or deviation of the domain from the external direction
doesn't change the steepness of the turn from ``on'' to ``off''
we only need to worry about the energy cost due to the sensors fighting
against the field and not against each other. So we need to understand
the activation energy 
\[
E=-\sum_{i}H_{i}S_{i}\rightarrow-\int d\phi N(\phi)S\left(\phi\right)H(\phi)
\]
where $N(\phi)$ is the number of receptors at the angle $\phi$ and $S\left(\phi\right)$ the average activity of receptors.  In practice the field $H\left(\phi\right)$ and the domain $S(\phi)$
point in arbitrary and distinct directions $\hat{\varphi}$ and $\hat{b}$
respectively. The energy cost as as a function
of the separation $\gamma=\hat{\varphi}-\hat{b}$ is given by 
\begin{eqnarray*}
E & = & -\int d\phi N\left(\phi-\hat{b}\right) S\left(\phi-\hat{b}\right)H(\phi-\hat{\varphi})\\
 & = & -pS_{0}\frac{N}{2}\int d\phi cos\left(\phi-\hat{\varphi} \right)sin\left(\phi-\hat{b} \right)tanh\left(Acos\left(\phi-\hat{b}\right)\right)\\
 & = & -pS_{0}\frac{N}{2}cos \gamma \int d\phi'cos\left(\phi'\right)sin\left(\phi'\right)tanh\left(Acos\left(\phi'\right)\right)\\
 & + & -pS_{0}\frac{N}{2}sin \gamma \int d\phi'sin\left(\phi'\right)sin\left(\phi'\right)tanh\left(Acos\left(\phi'\right)\right)\\
 & = & -pS_{0}\frac{N}{2}cos \gamma f(A)
\end{eqnarray*}
where we used $N(\phi)=\frac{N}{2}sin(\phi)$ and have set $f(A)=\int d\phi'cos\left(\phi'\right)sin\left(\phi'\right)tanh\left(Acos\left(\phi'\right)\right)$ .  This function is easy to integrate numerically and  is plotted below as a function of $A$.
  \begin{figure}[h]
\includegraphics[width=.8\linewidth,clip=]{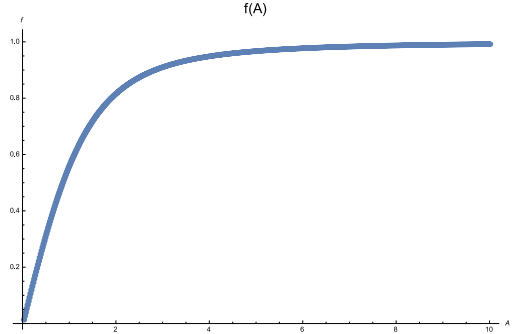}
 \caption{\label{fig:lattice} Integration constant $f(A)$ for "1/2-state"  }
 \end{figure}

In the idealized limit (which is very close to what happens in the
Ising model setup we have) $A>>1 \Rightarrow Tanh\left(Acos\left(\phi'\right)\right)\sim Step\; Function\;\Rightarrow f_{1}(A)\rightarrow1,\; S_{0}\rightarrow1$
. This is the idealized ``1/2-state'' and has energy given by
\[
E=-p\frac{N}{2}cos\gamma
\]
So we can see that when the domain is correctly aligned with the external
gradient ($\gamma=0$) the system sits in the bottom of an energy well
with depth $-p\frac{N}{2}$ and pushing the orientation away from
the external gradient takes energy to move up the wall. 

We would also like to know what the energy cost is for a non-zero
background field. We'll only do this for the idealized state where
the activity is perfectly divided by the point which the effective
field $H(\phi^{*})=0$ which happens at $\phi^{*}=cos^{-1}(\frac{H_{0}}{p})+\hat{\varphi}$ giving 
\begin{eqnarray*}
E & = & -\int d\phi N\left(\phi-\hat{b}\right) S\left(\phi-\hat{b}\right)H(\phi-\hat{\varphi})\\ & = & -\frac{N}{2}\int d\phi\left(pcos(\phi-\hat{\varphi})+H_{0}\right)sin\phi S\left(\phi\right)\\
 & = & -\frac{N}{2}\int_{2\pi-\phi^{*}+\gamma}^{\phi^{*}+\gamma}d\phi\left(pcos(\phi)+H_{0}\right)\\
 & + & \frac{N}{2}\int_{\phi^{*}+\gamma}^{2\pi-\phi^{*}+\gamma}d\phi\left(pcos(\phi)+H_{0}\right)\\
 & = & -p\frac{N}{2}\left(2\pi-4cos^{-1}(-\frac{H_{0}}{p})-4\sqrt{1-\frac{H_{0}^{2}}{p^{2}}}pcos\gamma \right)\\
\end{eqnarray*}
With the energy we can construct the probability of the "1/2-state" pointing away from the external direction as 
\begin{eqnarray*}
P(\gamma) & = & \frac{e^{-E(\gamma)}}{\int d\gamma e^{-E(\gamma)}}
 \end{eqnarray*}
Using the above derived energy and expanding $cos\gamma \sim1-\frac{1}{2}\gamma^{2}$
yields 
\[
P\left(\gamma \right)=\mathcal{N}(0,\sigma)
\]
with a variance given for the zero-field system $\sigma^{-2}=pNS_{0}f(A)/2$ which for the idealized zero constant field is  $\sigma^{-2}=pN/2$
and for the idealized non-zero constant field $\sigma^{-2}=p\sqrt{1-\frac{H_{0}^{2}}{p^{2}}}N/2$

\subsection{Monte-Carlo Information}

The computational results where made by simulating the behavior of $N=5,000$ sensors configured in a 2 dimensional network which has a spherical topology. The relaxation of the system was done using the Metropolis algorithm with the energies given by eq (1) of the paper.   Averages where done over 500 system realizations.

\end{document}